\RequirePackage{fix-cm}
\documentclass[smallextended,epjc3]{svjour3}  
\smartqed  
\RequirePackage{graphicx}
\journalname{Eur. Phys. J. C}
\usepackage{amsmath}
\usepackage{cite}
\usepackage[colorinlistoftodos]{todonotes}
\usepackage[colorlinks=true, allcolors=cyan]{hyperref}

\newcommand{\ben}{\begin{equation}}
\newcommand{\een}{\end{equation}}
\newcommand{\be}{\begin{equation*}}	
\newcommand{\ee}{\end{equation*}}
\newcommand{\bg}{\begin{gathered}}
\newcommand{\eg}{\end{gathered}}

\begin{document}

\title{Stability under scalar perturbations and quasinormal modes of 4D Einstein-Born-Infeld dilaton spacetime: Exact spectrum
}

\author{ 
			Kyriakos Destounis   		\thanksref{e1,addr1}
		\and
			Grigoris Panotopoulos   	\thanksref{e2,addr1}
		\and
			{\'A}ngel Rinc{\'o}n 		\thanksref{e3,addr2}
}

\thankstext{e1}{e-mail: \href{mailto:kyriakosdestounis@tecnico.ulisboa.pt}{\nolinkurl{kyriakosdestounis@tecnico.ulisboa.pt}}}
\thankstext{e2}{e-mail: \href{mailto:grigorios.panotopoulos@tecnico.ulisboa.pt}{\nolinkurl{grigorios.panotopoulos@tecnico.ulisboa.pt}}}
\thankstext{e3}{e-mail: \href{mailto:arrincon@uc.cl}{\nolinkurl{arrincon@uc.cl}}}


\institute{Centro de Astrof\'{\i}sica e Gravita{\c c}{\~a}o-CENTRA, Instituto Superior T{\'e}cnico-IST, 
\\
Universidade de Lisboa-UL, Avenida Rovisco Pais, 1049-001 Lisboa, Portugal.
\label{addr1}  \and
Instituto de F{\'i}sica, Pontificia Universidad Cat{\'o}lica de Chile,
\\
Av. Vicu{\~n}a Mackenna 4860, Santiago, Chile. \label{addr2}    
}

\date{Received: date / Accepted: date}

\maketitle

\begin{abstract}
We study the stability under scalar perturbations, and we compute the quasinormal modes of the Einstein-Born-Infeld dilaton spacetime in 1+3 dimensions. Solving the full radial equation in terms of hypergeometric functions, we provide an exact analytical expression for the spectrum. 
We find that the frequencies are purely imaginary, and we confirm our results by computing them numerically. Although the scalar field that perturbs the black hole is electrically neutral, an instability similar to that seen in charged scalar perturbations of the Reissner-Nordstr{\"o}m black hole is observed.

\end{abstract}

\section{Introduction}

Within the framework of Einstein's General Relativity (GR) \cite{GR} black holes (BHs) are predicted to exist, and they provide us with en excellent playground to 
study and understand several aspects of gravitational theories. They have become objects of paramount importance to classical and quantum gravity after Hawking's original computation \cite{hawking1,hawking2}, in which he showed that radiation was emitted from the horizon of the BHs. In particular, the study of BHs has received considerable attention in the context of scale dependent theories (see e.g \cite{Koch:2016uso,Rincon:2017ypd,Rincon:2017goj,Rincon:2017ayr,Contreras:2017eza}), where certain deviations from the classical solution appear.

On the other hand, how a system responds to small perturbations has always been an important issue in Physics. The work of \cite{wheeler} marked the birth of BH perturbations, and it was later extended by \cite{zerilli1,zerilli2,zerilli3,moncrief,teukolsky}. The state-of-the art in BH perturbations is summarized in Chandrasekhar's monograph \cite{monograph}. When BHs are perturbed the geometry of spacetime undergoes dumbed oscillations. Quasinormal modes (QNMs), with a non-vanishing imaginary part, carry unique information about the few BH parameters, since they do not depend on the initial conditions. 
The strongest evidence so far that BHs do exist comes from the historical LIGO direct detections of gravitational waves \cite{ligo1,ligo2,ligo3}, that have opened a new window to our Universe. Therefore nowadays the QNMs of black holes are more relevant than ever, since by observing the quasinormal frequencies and damping rates we may determine the mass, the angular momentum and the charges of the BH, or even falsify the theoretical paradigm of the no-hair conjecture \cite{misner}.
QNMs of BHs have been extensively studied in the literature. For a review on the subject see \cite{review1}, and for a more recent one \cite{review2}.

In the literature traditionally relativistic scattering of waves has been studied in asymptotically flat spacetimes, such as Schwarzschild \cite{SBH}, Kerr \cite{kerr} and Reissner-Norstrom \cite{RN} BHs, see e.g. \cite{col1,col2,col3,col4,col5,col6,col7}. However, due to inflation \cite{guth}, the AdS/CFT correspondence \cite{adscft1,adscft2}, and the current cosmic acceleration \cite{SN1,SN2}, asymptotically non-flat spacetimes with a non-vanishing cosmological constant have also been studied over the years \cite{kanti,kanti2,kanti3,3D1,3D2,Panotopoulos:2017yoe,
Panotopoulos:2016wuu,Fernando:2004ay,coupling,chinos,Ahmed:2016lou}.
In \cite{dilatonBI1,dilatonBI2}, however, the authors have found black hole solutions in three and four dimensions that are neither asymptotically flat nor asymptotically (anti) de Sitter. In those works the model is described by the Einstein-Born-Infeld dilaton action. Originally the Born-Infeld non-linear electrodynamics was introduced in the 30's to obtain a finite self-energy of point-like charges \cite{BI}. In more recent times this type of action reappears in the open sector of superstring theories \cite{ST1,ST2} as it describes the dynamics of D-branes \cite{Dbranes1,Dbranes2}. Furthermore, in the closed sector of all superstring theories at the massless level the graviton is accompanied by the dilaton that determines the string coupling constant. Since superstring theory is so far the only consistent theory of quantum gravity, it would be interesting to study the QNM of gravitational backgrounds obtained in the framework of Einstein-Born-Infeld dilaton models.

Computing the QNM frequencies in an analytical way is possible in a few cases only \cite{cardoso2,exact,potential,ferrari1,zanelli,fernando1,fernando2}, while in most of the cases some numerical scheme \cite{KaiLin1,KaiLin2,KaiLin3,Aron} or semi-analytical methods are employed, such as the well-known from standard quantum mechanics WKB method used extensively in the literature \cite{wkb1,wkb2,paper1,paper2,paper3,paper4,paper5,paper6,paper7,paper8,Panotopoulos:2017hns}. 
%
%
%
In the present work we obtain an exact analytical expression for the quasinormal spectrum of a four-dimensional Einstein-Born-Infeld dilaton spacetime, which is well motivated since it contains ingredients found in superstring theory. For a neutral BH, such as the Schwarzschild one, the scalar, vector and tensor perturbations can be studied separately. If, however, the BH is electrically charged then electromagnetic and gravitational perturbations are coupled and must be studied simultaneously \cite{paper2}. Therefore, in this work we take the first step to study the stability under scalar perturbations by perturbing the BH with a probe scalar field $\Phi$, not to be confused with the dilaton $\phi$ (see
the discussion below), hoping to be able to address the problem of the coupled electromagnetic-gravitational perturbation in a future work.

Our work is organized as follows: After this introduction, we present the model and the BH solution in the next section. In section 3 we discuss scalar perturbations where we present the effective potential of the Schr{\"o}dinger-like equation, while in the fourth section we solve the radial equation in terms of hypergeometric functions. In section 5 we obtain an exact expression for the quasinormal modes, and in section 6 we compare our solution with numerical results obtained with a recently developed method \cite{KaiLin1}.
Finally, we conclude our work in the last section. We use natural units such that $c = \hbar = 1$ and metric signature $(-, +, +, +)$.

\section{The model and the BH background}\label{solution}

We consider the model described by the action
\begin{align}
S[g_{\mu \nu},A_{\mu \nu},\phi] = \int \mathrm{d}^4 x \sqrt{-g} \Bigl[ &R-2(\nabla \phi)^2-V(\phi) \ +
\\
&4 \gamma e^{-2 \kappa \phi} (1-\sqrt{1+Y}) \Bigl],\nonumber
\end{align}
where
\begin{equation}
Y = \frac{F_{\mu \nu} F^{\mu \nu}}{2 \gamma},
\end{equation}
and where $R$ is the Ricci scalar, $g$ is the determinant of the metric tensor $g_{\mu \nu}$, $F_{\mu \nu}$ is the electromagnetic field strength, $\phi$ is the dilaton with a self-interaction potential $V(\phi)=2 \Lambda e^{-2 \kappa \phi}$, $\gamma$ is the Born-Infeld parameter, and $\kappa$ is the dilaton coupling constant. Assuming static spherically symmetric solutions, the line element of the metric is found to be \cite{dilatonBI2}
\begin{equation}
\mathrm{d} s^2=-f(r) \mathrm{d}t^2 + f(r)^{-1} \mathrm{d}r^2 + e^{2 \kappa \phi} (\mathrm{d} \theta^2 + \sin^2 \theta \: \mathrm{d}  \varphi ^2),
\end{equation}
while the dilaton is given by \cite{dilatonBI2}
\begin{equation}
\phi(r) = \frac{\kappa}{1+\kappa^2} \ln(br-c),
\end{equation}
where $b,c$ are constants of integration. In the following we set for convenience and without loss of generality $b=1$ and $c=0$.
Since the model is string inspired, in the following
we shall consider the string coupling case $\kappa=1$. Then 
the line element takes the form
\begin{equation}
\mathrm{d}s^2=- \left( \frac{r}{L} - r_0 \right) \mathrm{d}t^2 + \left( \frac{r}{L} - r_0 \right)^{-1} \mathrm{d}r^2 + r (\mathrm{d} \theta^2 + \sin^2 \theta \: \mathrm{d}  \varphi ^2),
\end{equation}
where the constant $r_0$ is related to the mass of the black hole, $r_0=4M$ \cite{dilatonBI2}, while $L$ is given by
\begin{equation}
L^{-1} = 2 (1-\Lambda-2 H),
\end{equation}
where the constant $H$ is given by \cite{dilatonBI2}
\begin{equation}
H = -\gamma + \sqrt{\gamma (Q^2+\gamma)},
\end{equation}
and the charge $Q$ of the black hole is given by \cite{dilatonBI2}
\begin{equation}
Q^2 = \frac{1+\sqrt{1+16 \gamma^2}}{8 \gamma}.
\end{equation}
There is a single event horizon $r_H=L r_0$, and therefore the metric function can be written down equivalently as $f(r)=(r-r_H)/L$. Overall, the model is characterized by 3 free parameters, namely $\gamma, \Lambda, M$. The horizon depends on all of them while, while $L$ does not depend on the mass of the black hole.

\section{Scalar perturbations}

\begin{figure}[ht!]
\centering
\includegraphics[scale=0.8]{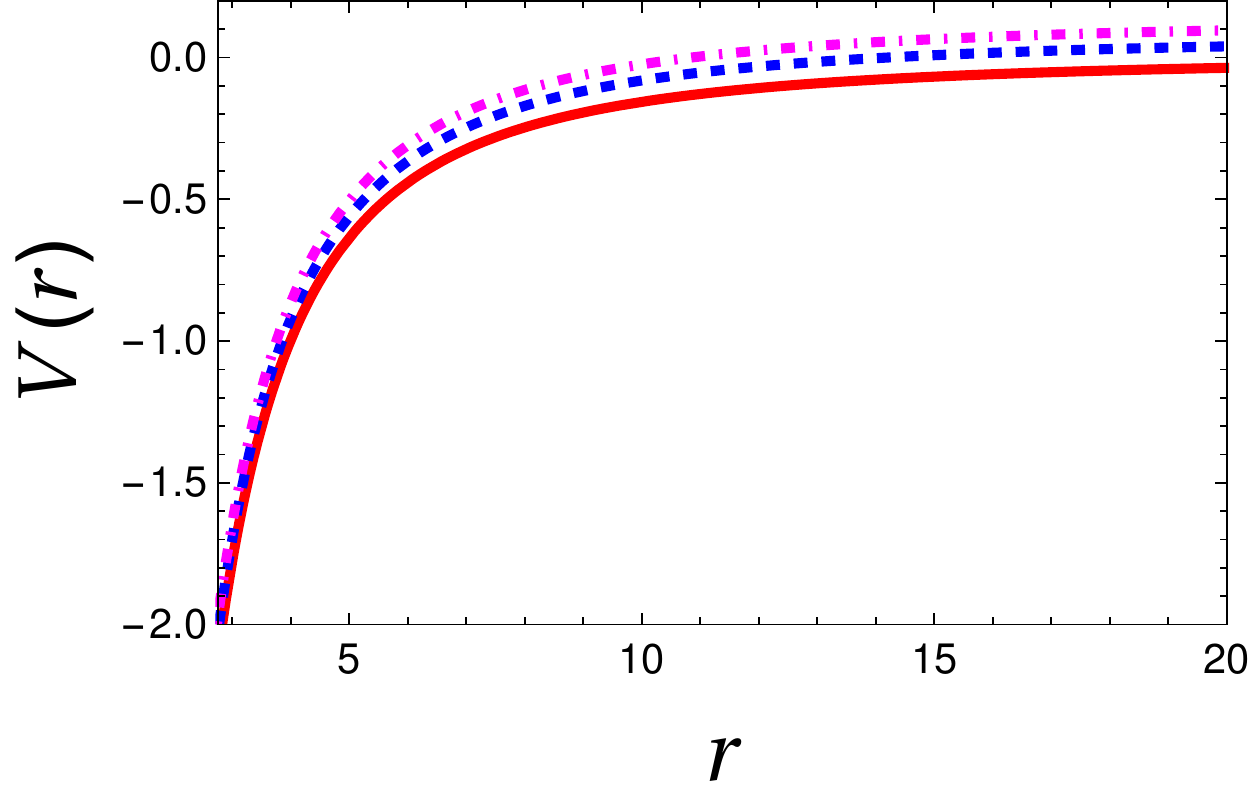}
\caption{\label{fig:1}
Effective potential versus $r$ for $l = 0, M=2, \Lambda=0.1$ and $\gamma=0.1$ (solid red line), $\gamma=0.5$ (dashed blue line) and $\gamma=2$ (dotted-dashed magenta line).
}
\end{figure}

\begin{figure}[ht!]
\centering
\includegraphics[scale=0.8]{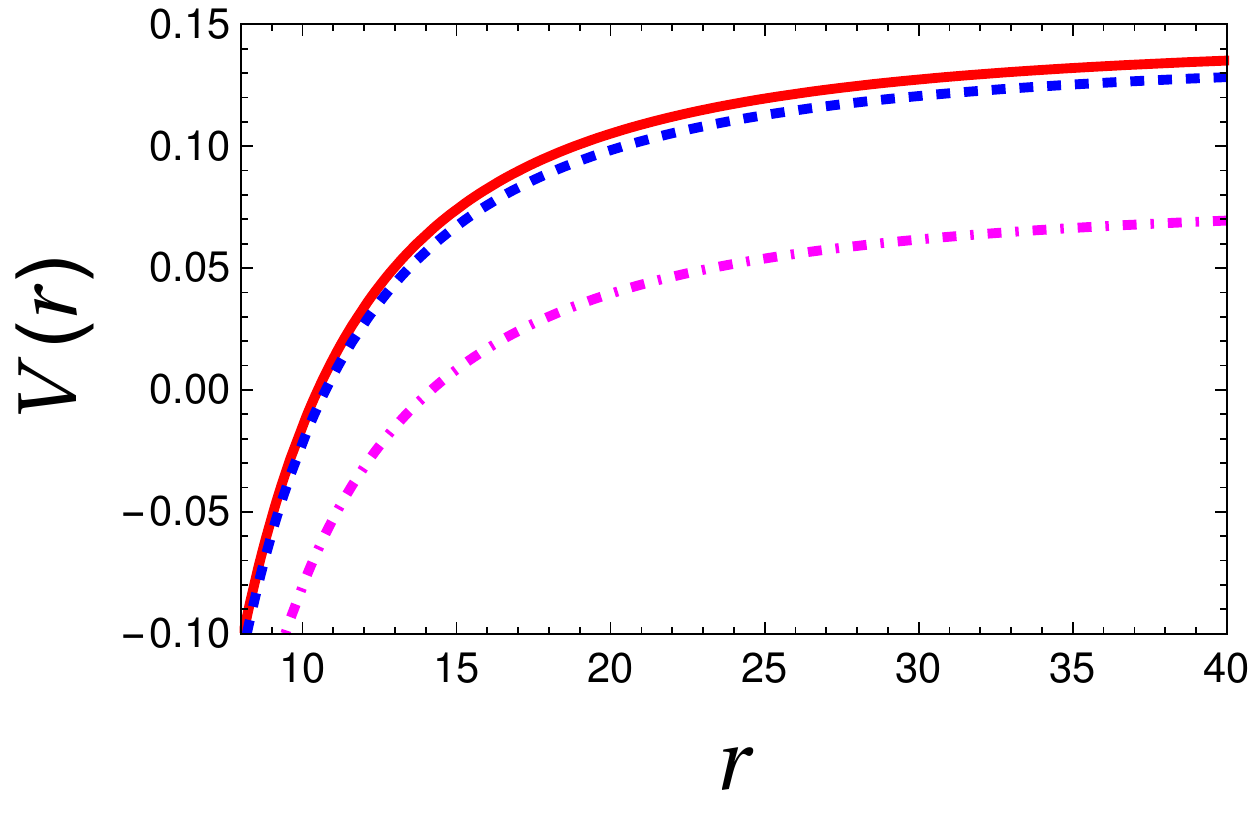}
\caption{\label{fig:2}
Effective potential versus $r$ for $l = 0, M=2, \gamma=0.5$ and $\Lambda=0.001$ (solid red line), $\Lambda=0.01$ (dashed blue line) and $\Lambda=0.1$ (dotted-dashed magenta line).
}
\end{figure}

In this section we study the propagation of a probe minimally coupled massless scalar field $\Phi(t,r,\theta, \varphi )$, not to be confused with the dilaton $\phi$,
in a given gravitational background of the form
\begin{equation}
\mathrm{d}s^2 = -h(r) \mathrm{d}t^2 + h(r)^{-1} \mathrm{d}r^2 + r (\mathrm{d} \theta^2 + \sin^2 \theta \: \mathrm{d}  \varphi ^2),
\end{equation}
with a known metric function $h(r)=(r-r_H)/L$. The starting point is the well-known wave equation
\begin{equation}
\frac{1}{\sqrt{-g}} \partial_\mu (\sqrt{-g} g^{\mu \nu} \partial_\nu) \Phi = 0,
\end{equation}
which is a partial differential equation for the scalar field. Next we seek solutions where the time and angular dependence are known as follows
\begin{equation}\label{separable}
\Phi(t,r,\theta, \varphi ) = e^{i \omega t} R(r) Y_l^m(\theta,  \varphi ),
\end{equation}
with $Y_l^m$ being the usual spherical harmonics. Using the above ansatz it is straightforward to obtain the radial equation, which is an ordinary differential equation
\begin{equation}\label{radial}
R'' + \left(\frac{h'}{h}+\frac{1}{r}\right) R' + \left(\frac{\omega^2}{h^2}-\frac{l (l+1)}{r h}\right) R = 0,
\end{equation}
where the prime denotes differentiation with respect to radial coordinate $r$. Next, we recast the equation for the radial part into a Schr{\"o}dinger-like equation of the form
\begin{equation}
\frac{\mathrm{d}^2 \psi}{\mathrm{d} x^2} + (\omega^2 - V(x)) \psi = 0,
\end{equation}
by defining new variables, a dependent $R \rightarrow \psi$ as well as an independent one $r \rightarrow x$ as follows
\begin{eqnarray}
R & = & \frac{\psi}{\sqrt{r}}, \\
x & = & \int \frac{\mathrm{d} r}{h(r)}=L \ln\Bigl(\frac{r-r_H}{d}\Bigl),
\end{eqnarray}
with $x$ being the so-called tortoise coordinate, and $d$ is a constant of integration which will be taken as unity. Therefore, we obtain the expression for the effective potential
\begin{equation}
V(r) = h(r) \: \left( \frac{l (l+1)}{r}+\frac{h'(r)}{2 r}-\frac{h(r)}{4 r^2} \right),
\end{equation}
which can be simplified to be
\begin{equation}
V(r) = V_0 - \frac{r_H l (l+1)}{L r} - \frac{r_H^2}{4 L^2 r^2},
\end{equation}
where the constant term is given by $V_0=(Ll(l+1)+1/4)/L^2$. The effective potential barrier versus the radial coordinate can be 
seen in figures \ref{fig:1} and \ref{fig:2} where we plot it for different $\gamma$ and $\Lambda$, respectively. Since it does not exhibit a maximum the WKB approximation is not applicable, and therefore we shall turn our attention to a different method for the numerical verification of our main analytical expression, see eq. (\ref{exact_QNM}) below.

To complete the formulation of the physical problem, we must also impose the appropriate boundary conditions at horizon and at infinity, which are the following \cite{superradiance}
\begin{equation}
\psi(x) \rightarrow
\left\{
\begin{array}{lcl}
A e^{i \omega x} & \mbox{ if } & x \rightarrow - \infty \\
&
&
\\
 C_+ e^{i k_{\infty} x} + C_- e^{-i k_{\infty} x} & \mbox{ if } & x \rightarrow + \infty
\end{array}
\right.
\end{equation}
where $A, C_+, C_-$ are arbitrary constants, and $k_{\infty}$ depends on the value of the effective potential at infinity. If both $C_-$ and $C_+$ are non-zero the procedure just described allows us to compute the so-called greybody factors, which show the modification of the original spectrum of Hawking radiation due to the effective potential barrier. In this case the frequencies are real and take continuous values. For an incomplete list see e.g. \cite{col1,col2,col3,col4,col5,col6,col7,kanti,kanti2,kanti3,3D1,3D2,Panotopoulos:2017yoe,
Panotopoulos:2016wuu,Fernando:2004ay,coupling,chinos,Ahmed:2016lou} and references therein. If, on the other hand, we require that the first coefficient of the second condition vanishes, i.e. $C_+ = 0$, we obtain an infinite set of discrete complex numbers $\omega_n=\omega_R + \omega_I i$, which are precisely the QNM frequencies of the black hole \cite{ferrari2}. The purely ingoing wave physically means that nothing can escape from the horizon, while the purely outgoing wave corresponds to the requirement that no radiation is incoming from infinity. Given the time dependence of the probe scalar field $\Phi \sim e^{i \omega t}$, it is clear that a positive imaginary part, $\omega_I > 0$, corresponds to stable modes, while a negative imaginary part, $\omega_I < 0$, corresponds to unstable modes. In the first case the real part of the mode $\omega_R$ determines the frequency of the oscillation, $\nu = \omega_R/(2 \pi)$, while the imaginary part $\omega_I$ describes the damping time, that is the decay of the fluctuation at a time scale $t_D=1/\omega_I$.

Since at the horizon the effective potential vanishes, the general solution for the function $\psi$ close to the horizon (where $\omega^2 \gg V(x)$)
is given by
\begin{equation}
\psi(x) = A_{+} e^{i \omega x} + A_{-} e^{-i \omega x},
\end{equation}
while requiring purely ingoing solution we set $A_{-}=0$ \cite{Fernando:2004ay,chinos}, and thus the solution becomes
\begin{equation}\label{ingoing}
\psi(x) = A e^{i \omega x}.
\end{equation}
On the other hand, it is easy to check that at large $r$ (or at large $x$, since when $r \gg r_H$, $r \simeq e^{x/L}$) the potential tends to
the constant $V_0$, and therefore
defining $\Omega \equiv \sqrt{\omega^2-V_0}$ the solution
for $\psi$ is given by
\begin{align}
\psi(x) &= D_{+} e^{i \Omega x} + D_{-} e^{-i \Omega x}.
\end{align}
Therefore, the far-field solution expressed in the tortoise coordinate $x$ takes the form of ingoing and outgoing plane waves provided that
$\omega^2 > V_0$, while the QNMs are determined by requiring that $D_+=0$, that is, the far-field solution takes the form
\begin{equation}\label{outgoing}
\psi(x)= D e^{-i \Omega x}.
\end{equation}

\section{Solution of the full radial equation in terms of hypergeometric functions}

Next, we find an exact solution of the radial equation (\ref{radial}) in terms of hypergeometric functions by introducing $z=1-r_H/r$. The new
equation for $z$ reads
\begin{equation}
z (1-z) R_{zz} + (1-z) R_z + \left( \frac{A}{z} + \frac{B}{-1+z} \right) R = 0,
\end{equation}
where $A=(\omega L)^2, B=-(\omega L)^2+L l (l+1)$. To get rid of the poles we set
\begin{equation}
R = z^\alpha (1-z)^\beta F,
\end{equation}
where now $F$ satisfies the following differential equation
\begin{multline}
z (1-z) F_{zz} + [1+2 \alpha - (1+2 \alpha+2 \beta) z] F_z
\\
+ \left( \frac{\bar{A}}{z} + \frac{\bar{B}}{-1+z} - C \right) F = 0,
\end{multline}
and the new constants are given by
\begin{eqnarray}
\bar{A} & = & A + \alpha^2, \\
\bar{B} & = & B + \beta - \beta^2, \\
C & = & (\alpha +\beta)^2.
\end{eqnarray}
Demanding that $\bar{A} = 0 = \bar{B}$ we obtain the
Gauss' hypergeometric equation
\begin{equation}
z (1-z) F_{zz} + [c-(1+a+b) z] F_z - ab F = 0,
\end{equation}
and we determine the parameters $\alpha, \beta$ as follows
\begin{eqnarray}
\alpha_\pm & = &\pm i \omega L, \\
\beta_\pm & = & \frac{1}{2} \pm i \sqrt{(\omega L)^2 - L l (l+1) -\frac{1}{4}} \label{beta}.
\end{eqnarray}
Finally, the three parameters of Gauss' equation are given by
\begin{eqnarray}
c & = & 1+2 \alpha, \\
a & = &  \alpha + \beta, \\
b & = & \alpha + \beta.
\end{eqnarray}
Note that the parameters $a,b,c$ satisfy the condition $c-a-b=1-2 \beta$.
Therefore, the general solution for the radial part is given by
\begin{equation}
\label{sol}
R(z) = z^\alpha (1-z)^\beta [ D_+ F(a,b;c;z) + D_- z^{1-c} F(a-c+1,b-c+1;2-c;z) ]
\end{equation}
where $D_-, D_+$ are arbitrary coefficients, and 
\begin{equation}
F(a,b;c;z) \,\,\,\,\,\,\,\,\,\,\,\, \text{and} \,\,\,\,\,\,\,\,\,\,\,\, z^{(1-c)}F(a-c+1,b-c+1;2-c;z).
\end{equation}
are the two linearly independent solutions of the Gauss' hypergeometric equation.

Close to the horizon, ($z \rightarrow 0$), the hypergeometric function $F(a,b;c;z)$ can be expanded in a Taylor series as follows
\begin{equation}
F(a,b;c;z) = 1 + \frac{a b}{c} \:z + ...
\end{equation}
Therefore the radial part becomes $R(z) \simeq D_+ z^\alpha+D_- z^{-\alpha}$,
and the solution (\ref{sol}) for the choice of $\alpha=\alpha_+$ reproduces the purely ingoing solution at the horizon (\ref{ingoing}) by setting $D_-=0$, where
the parameter $z$ can be written approximately $z \simeq (r-r_H)/r_H = e^{x/L}/r_H$. It is important to note that by choosing $\alpha=\alpha_-$ we can get (\ref{ingoing}) if we set $D_+=0$.

\section{Exact quasinormal spectrum}

\begin{figure}[ht!]
\centering
\includegraphics[scale=0.8]{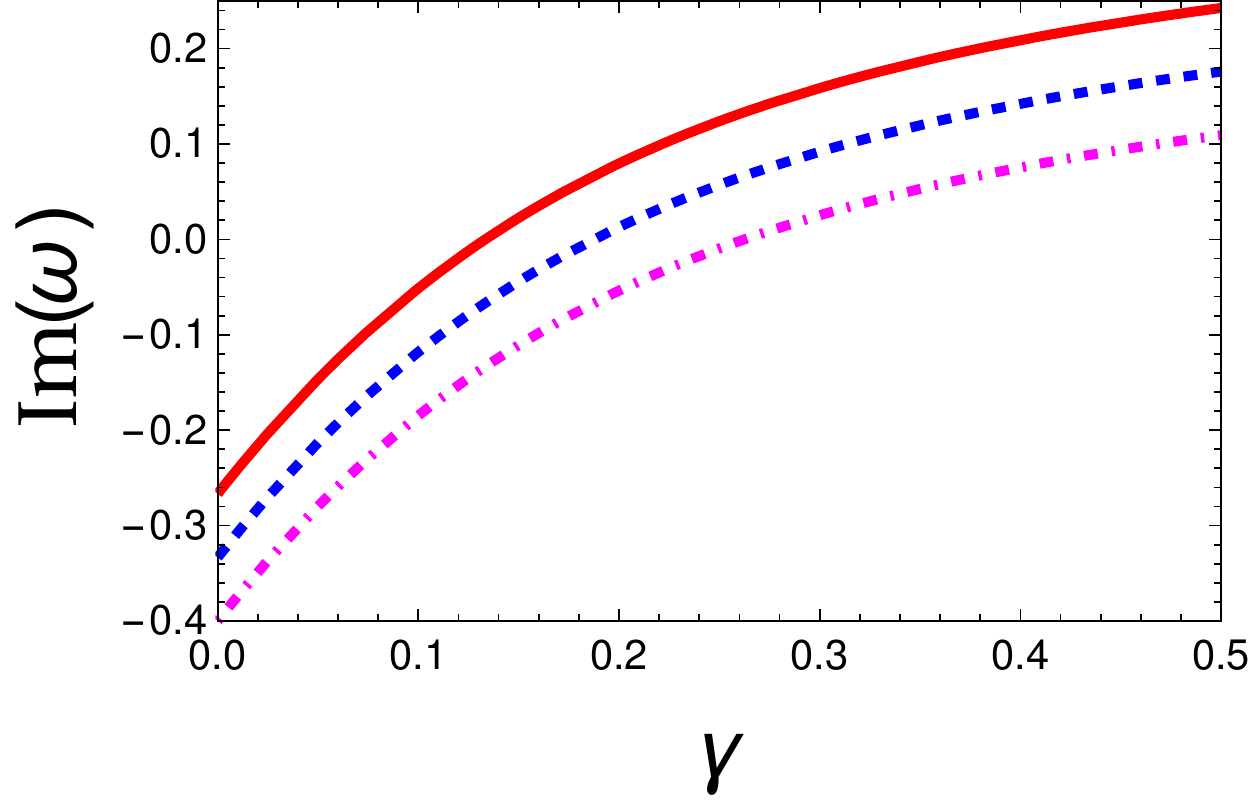}
\caption{\label{fig:3}
Imaginary part of the QN modes versus $\gamma$ for $l = 0, n=1$ and $\Lambda=0.2$ (solid red line), $\Lambda=0.25$ (dashed blue line) and $\Lambda=0.3$ (dotted-dashed magenta line).
}
\end{figure}

\begin{figure}[ht!]
\centering
\includegraphics[scale=0.8]{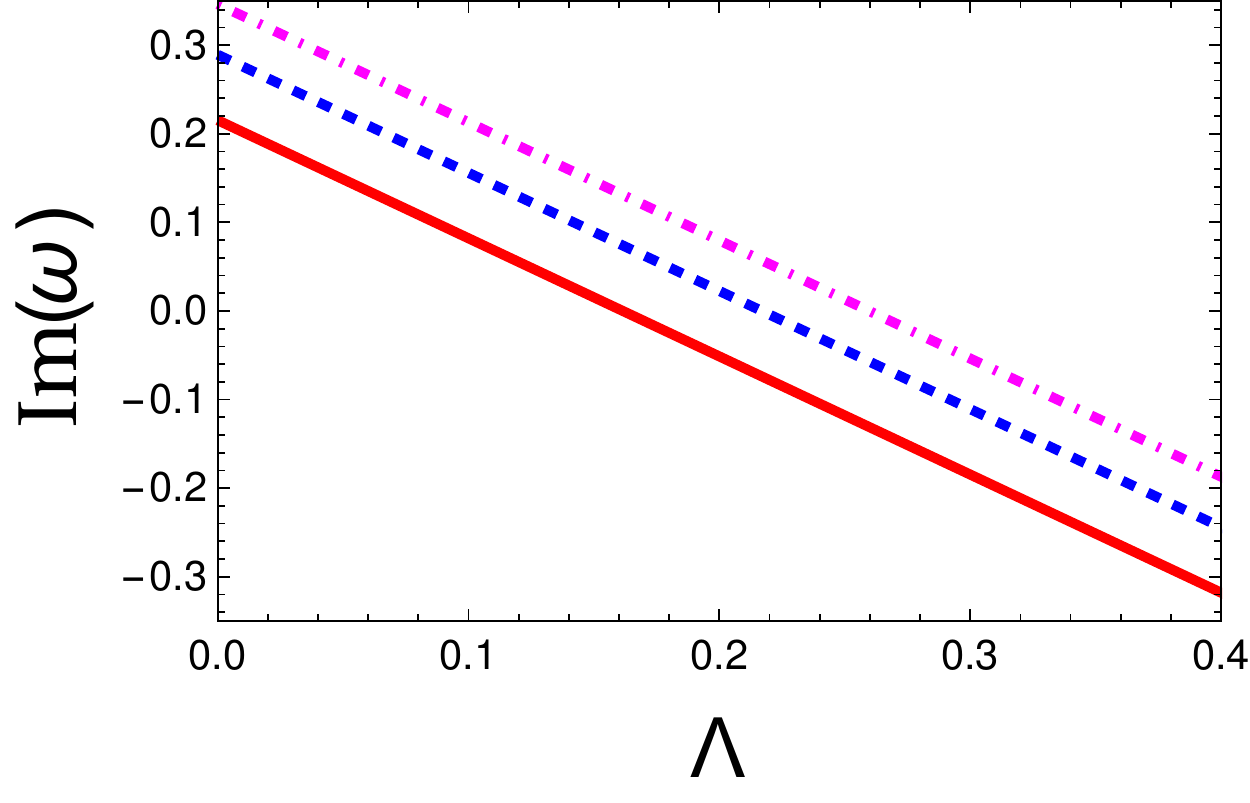}
\caption{\label{fig:4}
Imaginary part of the QN modes versus $\Lambda$ for $l = 0, n=1$ and $\gamma=0.1$ (solid red line), $\gamma=0.15$ (dashed blue line) and $\gamma=0.2$ (dotted-dashed magenta line).
}
\end{figure}

To see how the radial part behaves in the far-field zone $r \gg r_H$ (where $z \rightarrow 1$) we use the transformation \cite{handbook}
\begin{equation}
\begin{split}
F(a,b;c;z) = \ &\frac{\Gamma(c) \Gamma(c-a-b)}{\Gamma(c-a) \Gamma(c-b)}
\ \times
\\
&F(a,b;a+b-c+1;1-z) \ +
\\
 (1-z)^{c-a-b} &\frac{\Gamma(c) \Gamma(a+b-c)}{\Gamma(a) \Gamma(b)}
\ \times
\\
&F(c-a,c-b;c-a-b+1;1-z),
\end{split}
\end{equation}
and therefore the radial part as $z \rightarrow 1$ reads
\begin{equation}
\begin{split}
R(z \rightarrow 1)=D (1-z)^\beta \frac{\Gamma(1+2 \alpha) \Gamma(1-2 \beta)}{\Gamma(1+\alpha-\beta) \Gamma(1+\alpha-\beta)} \\
+D (1-z)^{1-\beta} \frac{\Gamma(1+2 \alpha) \Gamma(-1+2 \beta)}{\Gamma(\alpha+\beta) \Gamma(\alpha+\beta)}.
\end{split}
\end{equation}
Since $z=1-(r_H/r)$, the radial part $R(r)$ for $r \gg r_H$ can be written down as follows
\begin{equation}
\begin{split}
R(r) \simeq D \frac{\Gamma(1+2 \alpha) \Gamma(1-2 \beta)}{\Gamma(1+\alpha-\beta) \Gamma(1+\alpha-\beta)}
\left(\frac{r_H}{r}\right)^{\beta} \\
+D \frac{\Gamma(1+2 \alpha) \Gamma(-1+2 \beta)}{\Gamma(\alpha+\beta) \Gamma(\alpha+\beta)} \left(\frac{r_H}{r}\right)^{1-\beta}.
\end{split}
\end{equation}
Upon defining new constants $D_1, D_2$ as follows
\begin{eqnarray}
D_1 & = & D \: \frac{\Gamma(1+2 \alpha) \Gamma(1-2 \beta)}{\Gamma(1+\alpha-\beta) \Gamma(1+\alpha-\beta)}, \\
D_2 & = & D \: \frac{\Gamma(1+2 \alpha) \Gamma(-1+2 \beta)}{\Gamma(\alpha+\beta) \Gamma(\alpha+\beta)},
\end{eqnarray}
the function $\psi$ that satisfies the Schr{\"o}dinger-like equation takes the form
\begin{equation}
\psi \simeq D_1 \ r_H^\beta e^{-i  \text{Im}(\beta) \frac{x}{L}} + D_2 \ r_H^{1-\beta} e^{i  \text{Im}(\beta) \frac{x}{L}}.
\end{equation}
In the final step of the calculation, if we choose $\beta=\beta_+$, we obtain the outgoing boundary condition (\ref{outgoing}) by setting $D_2=0$, as
we mentioned when we discussed scalar perturbation in section 3.
We require that
the Gamma function in the denominator has a pole, and therefore the QNMs are determined imposing the condition
\begin{equation}
\alpha + \beta = -n,
\end{equation}
with $n=0,1,2,...$ being the overtone number. Using $\alpha=\alpha_+$ and $\beta=\beta_+$ we obtain the formula
\begin{equation}
\label{exact_QNM}
 \boxed{\omega_n = i \left( \frac{n+\frac{1}{2}}{2L} - \frac{1}{8L (n+\frac{1}{2})} - \frac{l (l+1)}{2 (n+\frac{1}{2})} \right)}
 \end{equation}
which is our main result in the present work. We can immediately observe the following three features of the spectrum, namely a) the QNMs are purely imaginary, b) they do not depend on $r_H$, so they depend on $\gamma$ and $\Lambda$ only, but not on the mass of the black hole, and c) all modes for $n=0$ become $\omega_{n=0}=-l (l+1) i$, and therefore they do not depend on any of the BH properties.
In particular, the fundamental mode $l=0,n=0$ is precisely zero, while for $l > 0$ all modes corresponding to $n=0$ are unstable. In table \ref{numerical_table} we compare our exact values with the ones computed numerically, while in the figures \ref{fig:3} and \ref{fig:4} we show how the imaginary part of the frequencies change with $\gamma$ and with $\Lambda$ respectively for $l=0, n=1$. Our figures show that the mode changes sign depending on the value of the cosmological constant $\Lambda$ as well as the Born-Infeld parameter $\gamma$ (or equivalently the electric charge $Q$). In particular, for low charge (large $\gamma$) the imaginary part is positive, whereas as the charge grows at a certain point the imaginary part becomes negative. Interestingly enough, a behaviour similar to that found in \cite{charged1,charged2} is observed, although the scalar field that perturbs the BH in the present work is not electrically charged. Contrary to these works, however, where it was found that all modes with $l >0$ were stable, our results show that for any value of the angular momentum there is a certain value of the overtone number after which the modes become stable. 


\section{Numerical Results}

We review a non grid-based interpolation scheme, proposed by Lin et al. \cite{KaiLin1}. This method makes use of data points in a small region of a query point to estimate its derivatives by employing Taylor expansion. The data points can be scattered, therefore they do not sit on a grid.  A key step of the method is to discretize the unknown eigenfunction in order to transform a differential equation and its boundary conditions into a homogeneous matrix equation. Based on the information about $N$ scattered data points, Taylor series are carried out for the unknown eigenfunction up to $N-$th order for each discretized point. The resulting homogeneous system of linear algebraic equations is solved for the eigenvalue. A huge advantage of this method is that the discretization of the wave function and its derivatives are made to be independent of any specific metric through coordinate transformation.

This method has been tested thoroughly for its accuracy and efficiency to various differential equation and eigenvalue problems in \cite{KaiLin1}, \cite{KaiLin2}, and \cite{KaiLin3}. The QNM results have been compared with WKB approximation \cite{wkb1,wkb2} (up to the sixth order), Horowitz-Hubeny method \cite{HHmethod}, and continued fraction method \cite{cfmethod} achieving very good precision.

We have applied the present method to compute the scalar QNMs of a four-dimensional Einstein-Born-Infeld dilaton black hole, and our numerical results are summarized in table \ref{numerical_table}.

\begin{table}[ht!]
\centering
\begin{tabular}{||c | c | c | c ||} 
\hline
  \multicolumn{4}{||c||}{$\gamma=0.112, \, \Lambda=0.01$ ($L$=3.00993)} \\
   \hline
    $n$ & $l=0$ & $l=1$ & $l=2$\\ [0.5ex] 
    \hline
   0  & 0  & -2i & -6i \\ 
    & (0.000) & (-2.000i) & (-6.000i) \\
   \hline
   1 & 0.221489i & -0.445178i & -1.77851i \\
    & (0.221489i) & (-0.445178i) & (-1.77851i) \\  
   \hline
   2 & 0.39868i & -0.00131968i & -0.80132i \\  
    & (0.39868i) & (-0.00131968i) & (-0.80132i)  \\
    \hline
   3 & 0.569543i & 0.283829i & -0.2876i \\  
     & (0.569543i) & (0.283829i) & (-0.2876i)\\
   \hline\hline
\multicolumn{4}{||c||}{$\gamma=0.5, \, \Lambda=0.1$ ($L$=1.77326)} \\
 \hline
     $n$ & $l=0$ & $l=1$ & $l=2$\\ [0.5ex] 
     \hline
  0  &  0  & -2i &  -6i   \\ 
  & (0.000) & (-2.000i) & (-6.000i) \\
  \hline
 1 & 0.375955i & -0.290712i &  -1.62405i \\ 
 & (0.375955i) & (-0.290712i) & (-1.62405i) \\
 \hline
 2 & 0.676718i & 0.276718i & -0.523282i \\ 
  & (0.676718i) & (0.276718i) & (-0.523282i) \\
   \hline
   3 & 0.966741i & 0.681026i & 0.109598i \\ 
    & (0.966741i) & (0.681026i) & (0.109598i) \\
  \hline\hline
  \multicolumn{4}{||c||}{$\gamma=2, \, \Lambda=0.001$ ($L$=1.06867)} \\
 \hline
     $n$ & $l=0$ & $l=1$ & $l=2$\\ [0.5ex] 
     \hline
 0  & 0  & -2i & -6i \\ 
  & (0.000) & (-2.000i) & (-6.000i) \\
 \hline
 1 & 0.623828i & -0.0428385i & -1.37617i \\
  & (0.623828i) & (-0.0428385i) & (-1.37617i) \\  
 \hline
 2 & 1.12289i & 0.722891i & -0.0771093i \\  
  & (1.12289i) & (0.722891i) & (-0.0771093i) \\
  \hline
 3 & 1.60413i & 1.31842i & 0.746987i \\ 
   & (1.60413i) & (1.31842i) & (0.746987i) \\
   \hline
\end{tabular}

\caption{Scalar QNMs of Einstein-Born-Infeld dilaton black hole for various values of $\gamma$ and $\Lambda$. $l,\,n$ are the angular momentum and overtone number, respectively, and $L^{-1}=2(1-\Lambda-2H)$, see text. The values without the parenthesis are the exact QNMs, while the ones in the parenthesis are the numerical values.}
\label{numerical_table}
\end{table}

We observe that the numerical results agree perfectly with our main result shown in eq. (\ref{exact_QNM}). We immediately see that for $n=0$ the modes depend only on the angular momentum $l$, irrespectively of the choice of $\gamma$ and $\Lambda$, which is in agreement with $\omega_{n=0}=-l (l+1) i$ obtained in the previous section. For $l=0$, the aforementioned fundamental mode is exactly zero, while for $l\ge 1$, as $l$ increases more unstable modes begin to appear. Finally, we observe that as $L$ decreases the stable modes decay faster while the unstable ones grow slower.

\section{Conclusions}

To summarize, in this article we have studied the stability under scalar perturbations of (1+3)-dimensional Einstein-Born-Infeld dilaton spacetimes, 
and we have provided an exact analytical expression for the frequencies, which are found to be purely imaginary. We have confirmed our results computing the frequencies numerically using a recently developed non grid-based numerical scheme. In addition, an instability similar to that seen in charged scalar perturbations of the Reissner-Nordstr{\"o}m black hole is observed, although in our work the scalar field that perturbs the BH is not electrically charged.


\section*{Acknowlegements}

K. D. acknowledges financial support provided under the European Union's H2020 ERC Consolidator Grant "Matter and strong field gravity: New frontiers in Einstein's theory" grant agreement no. MaGRaTh-646597. G. P. thanks the Funda\c c\~ao para a Ci\^encia e Tecnologia (FCT), Portugal, for the financial support to the Center for Astrophysics and Gravitation-CENTRA, Instituto Superior T\'ecnico, Universidade de Lisboa, through the Grant No. UID/FIS/00099/2013. A.R. was supported by the CONICYT-PCHA/\- Doctorado Nacional/2015-21151658. 


\end{document}